\documentclass{PoS}
\usepackage{eurosym}
\usepackage{amsmath}

\title{The International Linear Collider}

\ShortTitle{The International Linear Collider}

\author{\speaker{Karsten Buesser}\\
        DESY, Notkestrasse 85, 22607 Hamburg, Germany\\
        E-mail: \email{karsten.buesser@desy.de}}

\abstract{The International Linear Collider (ILC) is a proposed electron-positron collider for the centre-of-mass energy range of 200 to 500 GeV and with upgrade options towards 1~TeV. The ILC would be the ideal tool to explore with high precision the properties of the new Higgs-like particle that has recently been discovered at the LHC with a mass of around 125~GeV. The ILC accelerator design is based on the mature superconducting technology that has been developed in the TESLA collaboration and that is currently being used for the European XFEL. The exploitation of the huge physics potential of the ILC is a challenge for the design of the ILC detectors.}

\FullConference{Proceedings of the Corfu Summer Institute 2012 "School and Workshops on Elementary Particle Physics and Gravity"\\
		September 8-27, 2012\\
		Corfu, Greece}

\begin{document}

\section{Introduction}

The recent discovery of the new Higgs-like boson at the Large Hadron Collider (LHC) opens the window into a new regime of particle physics, the Terascale. The measurement of the properties of this new particle will answer the question whether the Higgs mechanism as predicted by the Standard Model is responsible for the generation of masses. The LHC will continue to explore the high energy region for hints of possible more fundamental theories of nature. 

The LHC collides protons, which allows to reach high collision energies with reasonable efforts. However, in proton collisions the initial states depend on the dynamics of the quarks and gluons and are therefore not well known. Since many years, lepton colliders and hadron colliders are being used together to explore the high energy frontier. In an electron-positron collider, the conditions and quantum numbers of the initial states are very well known. The clean event signatures and topologies allow for precision measurements that have a predictive power well beyond the nominal collision energy reach.

The International Linear Collider is a proposed e$^+$e$^-$ collider for an energy range of E$_{cm}$~=~200~--~500~GeV with an upgrade path towards 1~TeV~\cite{BrauJames:2007aa, Elsen:2011zz}. The recently found new particle could be measured with highest precision at the ILC. In addition, potential New Physics could be discovered and explored either in direct measurements or by indirect extrapolations to highest energies.

\section{The Physics Programme of the ILC}

The properties of the Standard Model Higgs particle are theoretically completely known. It is therefore straight-forward to determine, whether the newly discovered boson is the single Higgs particle of the Standard Model, or if it is rather the first hint of something unknown and opens the door to a more fundamental description of the world. The key to the determination of the Higgs properties are precision measurements that can be done best at a lepton collider.

At the ILC, essentially all quantum numbers of the Higgs can be pinned down. Many properties like the Higgs mass can be measured in a model-independent way. Even hypothetical Higgs particles that decay invisibly are not only in the reach of discovery, but some of their properties can be determined with precision. The Higgs mechanism for the generation of mass is a central ingredient of the Standard Model. The key measurements for its establishment are the decay ratios of the Higgs into the Standard Model fermions. The ILC can determine the couplings of the Higgs to the fermions with high precision by measuring the decay ratios. Those couplings should be proportional to the masses of the particles if the Higgs mechanism is correct. Also the coupling of the Higgs to itself can be measured at the ILC if enough data can be collected. All these measurements, the mass, the quantum numbers, and the couplings, explore the Higgs sector of the Standard Model completely. By doing this with highest precision, these measurements open the door to mass scales that are way beyond the collision energies.

There are good arguments to believe, that the Standard Model is only a low energy approximation of a more fundamental theory. Popular extensions of the Standard Model like Supersymmetry (SUSY) predict a plethora of new particles that would need to be measured with precision. Though the LHC has not seen any indication for supersymmetric particles yet, the ILC can explore SUSY sectors that are not in reach of the LHC. In addition, the ILC will be a $t\bar{}t$ factory that will produce large samples of this particle and allows for top physics studies with unprecedented precision.

Following the progress of the LHC running, a staged scenario for the ILC physics programme is being discussed. In a first stage, at centre-of-mass energies of around 250~GeV, the ILC would be a Higgs factory, where the Higgs quantum number, its mass, and most of its couplings would be measured. In a second stage, the ILC could run at the $t\bar{t}$ threshold as a top factory. In the final stage of the baseline scenario, the ILC would run at E$_{cm}$=500~GeV for the measurement of the top-Higgs Yukawa coupling, the tri-linear couplings (HWW, ZHH) and finally the Higgs self coupling.

The ILC has a clear upgrade path towards collision energies of 1~TeV. If the results from the LHC runs at 14~TeV give indications of interesting physics, it can be decided whether an ILC upgrade or rather a multi-TeV lepton collider as CLIC~\cite{Lebrun:2012hj} would be the way forward. 

\section{The Accelerator Challenge}

Designing a linear collider for the Terascale is an unprecedented technical challenge where significant R\&D work has been spent on in the last decades. Building on the experience from SLC at SLAC, many international collaborations worked on designs for normal- and superconducting linear accelerators. The ILC, that has been born out of these efforts, is the first truly global accelerator project where all major accelerator labs around the world join forces in the Global Design Effort (GDE) for the technical design of the ILC.  

\subsection{Lepton Collider Scaling}
The lepton collider with the highest collision energies so far was LEP at CERN, where electron and positrons were accelerated in the 27~km long tunnel that now hosts the LHC. The maximum collision energies (E$_{cm}$) at LEP were at 209~GeV, so seemingly not too far away from the envisaged ILC energy range. Scaling circular lepton colliders to high energies is however limited by the synchrotron radiation losses that increase proportional to the fourth power of the beam energy. This effect can somewhat be mitigated by increasing the radius $r$ and therefore the length of the rings as the radiated power scales also with $1/r$. In the end a cost optimised solution needs to be found. While the cost for the acceleration power (\euro$_{RF}$) scales as 
$\text{\euro}_{RF}\sim E^4/r$ there is also a linear cost component that comes for the cost of the tunnelling, and the machine equipment: $\text{\euro}_{lin}\sim r$.
The total cost optimum for a storage ring $\text{\euro}_{SR} = \text{\euro}_{RF} + \text{\euro}_{lin}$ has an optimal radius $r_{opt} \sim E^2$
and scales as $\text{\euro}_{SR} \sim E^2$~\cite{Richter:1976ug}. Thus, building and operating lepton storage rings at very high energies becomes economically inefficient.

In a linear collider, the particles are not stored, but only accelerated once towards the collision point. The cost for the acceleration power scales with the length of the linacs and therefore with the beam energy: $\text{\euro}_{LC} \sim E$. It is obvious, that the linear collider concept becomes the more efficient and economic choice above a certain energy range. However, it should be noted that these scaling laws are only very rough and the exact choice of the accelerator scheme depends on the physics requirements like energy range and luminosity as well as on the available technologies. For high-luminosity high-energy lepton colliders, the transition region is closely above of the LEP collision energies. Though one could imagine to build a collider that reaches the threshold for production of a light Higgs boson as a storage ring~\cite{Blondel:2012ey}, is is clear that energies of 500~GeV (cms) or above can realistically only be reached in linear colliders. And most importantly, linear colliders offer convenient possibilities for energy-staged projects. It seems feasible to start the physics programme with a reduced energy and then, by adding more accelerating structures (and tunnel) the energy could easily be increased as there are no fundamental radiation limits as in storage rings.

\subsection{The Luminosity Issue}

One key parameter of any high energy lepton collider is the luminosity. As the total e$^+$e$^-$ cross section drops proportional to $1/E_{cm}^2$, the luminosity of the collider needs to be higher the higher the envisaged collision energy is. For a collider with Gaussian beam shapes, the luminosity is given as

\begin{equation} \label{eq:lumi}
\mathcal{L} = \frac{n_b N^2 f_{rep}}{4\pi \sigma_x \sigma_y} H_D,
\end{equation}

where $n_b$ denotes the number of particle bunches in one pulse (bunch train), N is the number of particles per bunch, $f_{rep}$ is the pulse repetition frequency, and $\sigma_{x,y}$ are the horizontal and vertical beam sizes. $H_D$ is an additional enhancement factor that comes from the intense beam-beam interactions (see below). Introducing the beam power $P_{beam} = n_b N f_{rep} E_{cm}$ and the RF-power $P_{RF} = P_{beam}/\eta_{RF}$ yields

\begin{equation}
\mathcal{L} = \frac{\eta_{RF} P_{RF} N}{4\pi \sigma_x \sigma_y E_{cm}} H_D
\end{equation}

where $\eta_{RF}$ is the conversion efficiency from the power of the accelerating radio frequency cavities to the beam. The ILC is designed for a luminosity of 1.8 $\times$ 10$^{34}$ cm$^{-2}$s$^{-1}$ at a collision energy of 500~GeV (c.f.\ table~\ref{table:ILCparameters}). Thus, with $f_{rep}$ = 5~Hz, $n_b$ =1312, and N = 2 $\times$10$^{10}$, the beam power (for both beams) yields to P$_{beam}$ = 10.5 MW. Assuming a total efficiency of $\approx$10\% for the conversion of the wall-plug (AC) power to the RF power and finally to the beam, yields an AC power of more than 100~MW just to accelerate the beams and maintain luminosity.

\begin{table}[tb] 
 \caption[ILC baseline parameters summary]{Summary table of the baseline parameters for the ILC for 230 and 500~GeV (cms) energies~\cite{ILC-tech}.}
\small
\begin{tabular}{ l l l r r } \hline 
Centre-of-mass energy & $E_{CM}$ & GeV & 230 & 500 \\  
\hline
Luminosity pulse repetition rate & & Hz & 5 & 5 \\ 
Bunch population & $N$ & $\times10^{10}$ & 2 & 2 \\ 
Number of bunches & $n_b$ & & 1312 & 1312 \\
Linac bunch interval & $\Delta t_b$ & ns & 554& 554 \\
RMS bunch length & $\sigma_z$ & $\mu$m & 300 & 300 \\ 
RMS horizontal beam size at IP & $\sigma^*_x$ & nm & 789 & 474 \\ 
RMS vertical beam size at IP & $\sigma^*_y$ & nm & 7.7 & 5.9 \\ 
Vertical disruption parameter & $D_y$ & & 24.5 & 24.6 \\ 
Fractional RMS energy loss to beamstrahlung & $\delta_{BS}$ & \% & 0.83 & 4.5 \\ 
Luminosity & $\cal(L)$ & $\times 10^{34}$ cm$^{-2}$ s$^{-1}$ & 0.67 & 1.8 \\
Fraction of $\cal{L}$ in top 1\% $E_{CM}$ & $\cal{L}_{0.01}$ & \% & 89 & 58 \\ 
\hline
\end{tabular}
\label{table:ILCparameters}
\end{table}

While the repetition frequency in a storage ring is large, e.g.\ 44 kHz at LEP, it is limited in a linear collider for various reasons. The preparation of the bunch trains in the damping rings and the RF filling times of the accelerating cavities allow for typical repetition frequencies of a few Hz. So intrinsically the luminosity is a factor of $\approx$1000 smaller in a linear collider. The luminosity can be recovered by focusing the beams to much smaller sizes in the collisions. While the beam spot size at LEP was in the order of 130$\times$6~$\mu$m$^2$, it will be 474$\times$5.9 nm$^2$ at the ILC. While this brings the luminosity up significantly, the large electrical fields of the squeezed bunches increase the beam-beam interactions. The resulting effect on the outgoing bunches can be afforded at a linear collider where the bunches are not reused, but disposed off in the beam dumps after the collisions. 

The beam-beam interactions have two relevant effects. While the particles of the bunches are influenced by the fields of the oppositely charged particles from the oncoming bunch, both bunches are focused onto each other. This results in typical additional luminosity enhancement factors of $H_D\approx2$. At the same time, the beam-beam interaction is the source of photon radiation. While the high energetic particles are focused onto the other bunch, they emit so called `beamstrahlung' photons. This radiation dilutes the luminosity spectrum of the beams. Typically, for the ILC at 500~GeV collision energy, only 58\% of the collisions happen at cms energies that are within the top 1\% fraction of the collision energy spectrum (c.f. table~\ref{table:ILCparameters}). While the beamstrahlung photons are emitted into the very forward directions, they can collide themselves during the bunch interactions and produce lower energetic background particles with larger transverse momentum that can be a source of backgrounds for the experiments.

The energy that is lost due to beamstrahlung production $\delta_{bs}$ can be parametrised by

\begin{equation} \label{eq:bs}
\deltaÐ{bs} \approx 0.86 \frac{r_e^3 N^2 \gamma}{\sigma_z(\sigma_x + \sigma_y)^2}
\end{equation}

where $r_e$ is the classical electron radius, $\gamma$ is the relativistic factor $E_{beam}/m_0c^2$, and $\sigma_z$ is the bunch length. Comparing equations~\ref{eq:lumi} and \ref{eq:bs} shows that the choice of the beams with very flat cross sections maximizes the luminosity while the beamstrahlung energy losses are kept at reasonable levels.

\subsection{The ILC Technical Design}
\begin{figure}[t]
\centerline{\includegraphics[width=.9\textwidth]{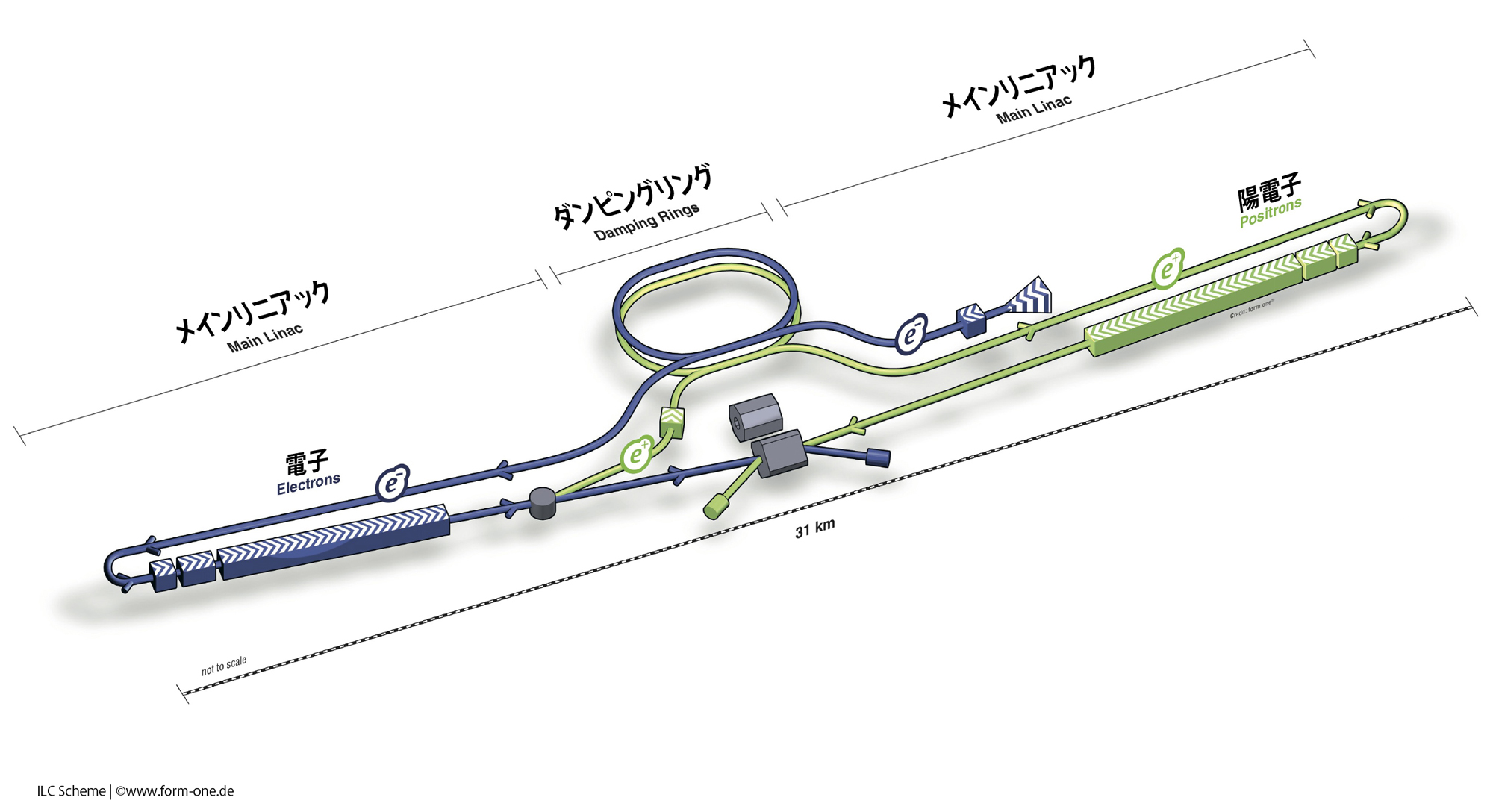}}
\caption{Schematic overview of the ILC layout. Figure: ILC GDE.}
\label{fig:ILC_outline}
\end{figure}

The ILC reference design has been published in the Reference Design Report~\cite{BrauJames:2007aa}, the publication of the Technical Design Report is envisaged for 2013. Figure~\ref{fig:ILC_outline} shows a schematic view of the ILC technical systems:
\begin{description}

\item[Electron Source] The electron source is a laser driven photo injector, where circular polarised photons illuminate a GaAs cathode. Longitudinal polarised electrons from the cathode are accelerated with high electric fields to 140-160 keV energies. Bunches of 1~ns length are produced at a rate of 3~MHz with peak currents of 5~nC/ns. The electron bunches have 80\% polarisation and are pre-accelerated to 5~GeV beam energy before they are injected into the damping rings.
\item[Positron Source] The positrons are produced with the high energy electron beam. The positron source is therefore located at the end of the electron linac. The 250~GeV energy electron beam is guided through a superconducting helical undulator. 30~MeV circular polarised photons are extracted towards a thin (0.4 radiation lengths) rotating target and produce e$^\pm$~pairs. The positrons are selected and focused in capture devices and also pre-accelerated to 5~GeV. The use of a thin target reduces multiple scattering and results in a better emittance of the positron beam. The positron bunches are polarised in the baseline version of the source to 30-40\%. 
\item[Damping Rings] The pre-accelerated electron and positron beams have emittances that are orders of magnitude too big to reach the small beam spot sizes in the collisions. Therefore the beams are stored for some time in the damping rings. Superconducting wigglers force the radiation of photons in a cone with the opening angle of $1/\gamma$ around the beam direction. At the same time, the accelerating structures in the damping rings recover the longitudinal momentum of the electrons and positrons. This interplay of radiation and acceleration reduces the emittance of the bunches. The damping time in the rings needs to be shorter than the time between two pulses in the linacs, so 200~ms in the 5~Hz operation mode (c.f.~table~\ref{table:ILCparameters}). Both rings, the one for the electrons and the one for the positrons, are housed in one single tunnel of 3.2~km circumference. The fact that a whole 1~ms long bunch train needs to be stored in each damping ring requires a compression on injection and a decompression on extraction by roughly a factor of $\approx$90. This sets stringent requirements to the injection and extraction kickers as the time separation of two bunches in the damping ring is only 8~ns.
\item[Main Linacs] After the extraction from the damping rings, the beams are transported to the end of the main linacs where they turn back and get accelerated to the collision energies. Beam acceleration in the 11~km long linacs is done using $\approx$7,400 superconducting niobium cavities~(c.f.~figure~\ref{fig:ILC_cavity}) with an average accelerating gradient of 31.5~MV/m and quality factors of $Q_0>10^{10}$.
\item[Beam Delivery Systems] The beam delivery system (BDS) transports the beams from the end of the main linacs to the interaction region. The main tasks of the BDS are the diagnostics of the beams after they leave the main linacs and the demagnification to the small beam spot sizes at the collision point. In addition, the beam halo needs to be removed in the collimation section to reduce the backgrounds in the detector. After the collision, the beams are transported to the main beam dumps that are designed for the 500~GeV beams of the ILC upgrade scenario, where 18~MW beam power need to be disposed in each dump.
\end{description}

\subsection{The Superconducting Technology}
\begin{figure}[t]
\centerline{\includegraphics[width=.6\textwidth]{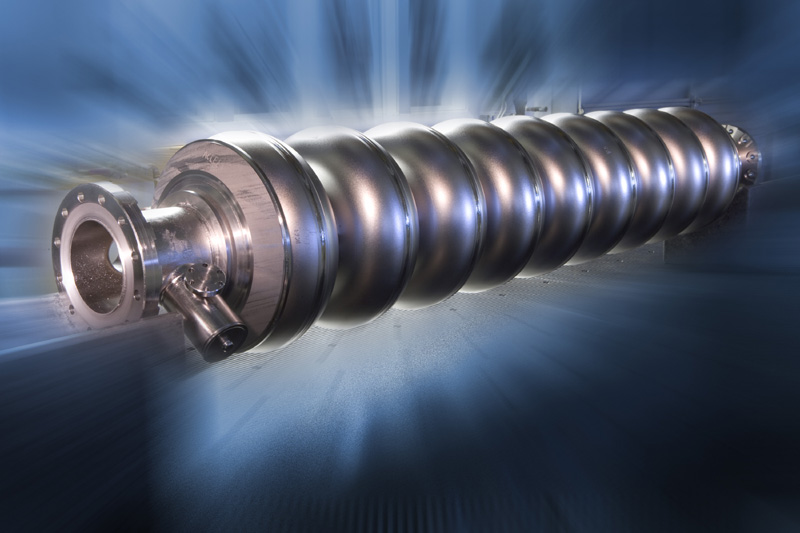}}
\caption{9-cell niobium accelerating cavity for the ILC. Photo: FNAL.}
\label{fig:ILC_cavity}
\end{figure}

The technological key to the ILC are the superconducting radio frequency cavities~(figure~\ref{fig:ILC_cavity}). These cavities operate at frequencies of 1.3~GHz and are cooled with 2~K liquid helium. The accelerating gradients are required to be on average at 31.5~MV/m with production yields of better than 80\%. The development of this superconducting technology goes back to the work of the TESLA collaboration~\cite{Aune:2000gb}. The same type of cavities are currently being produced for the European XFEL x-ray laser facility at DESY. The industrial production of 800 cavities for this 17.5~GeV electron linac is the ideal test case for the extrapolation to the production planning of the almost 16,000 cavities that are needed for the ILC.

\begin{figure}[t]
\centerline{\includegraphics[width=.7\textwidth]{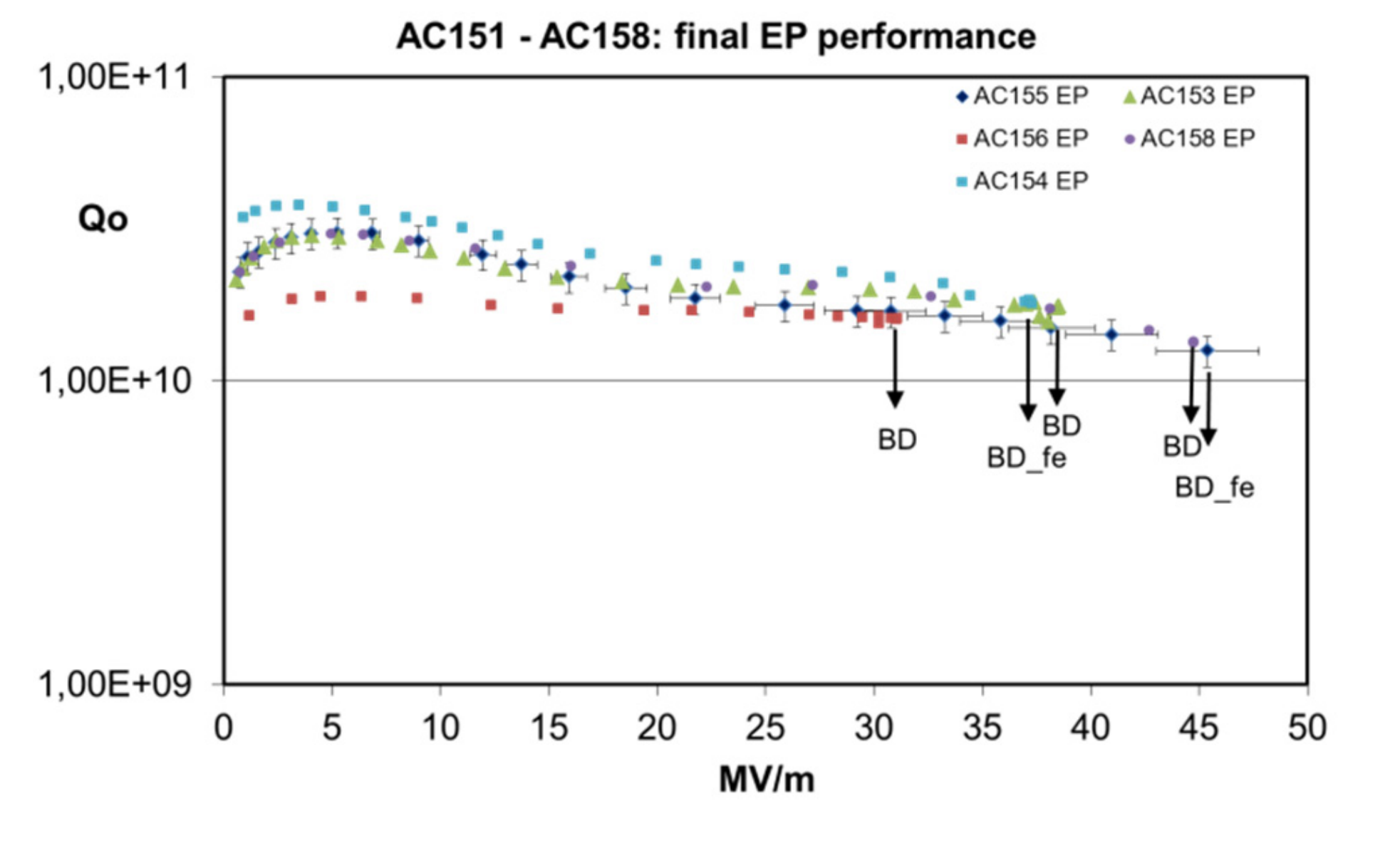}}
\caption{Quality factor $Q_0$ as a function of the accelerating gradient for ILC cavities~\cite{Reschke}.}
\label{fig:ILC_cavity_gradient}
\end{figure}

\begin{figure}[hbt]
\centerline{\includegraphics[width=.6\textwidth]{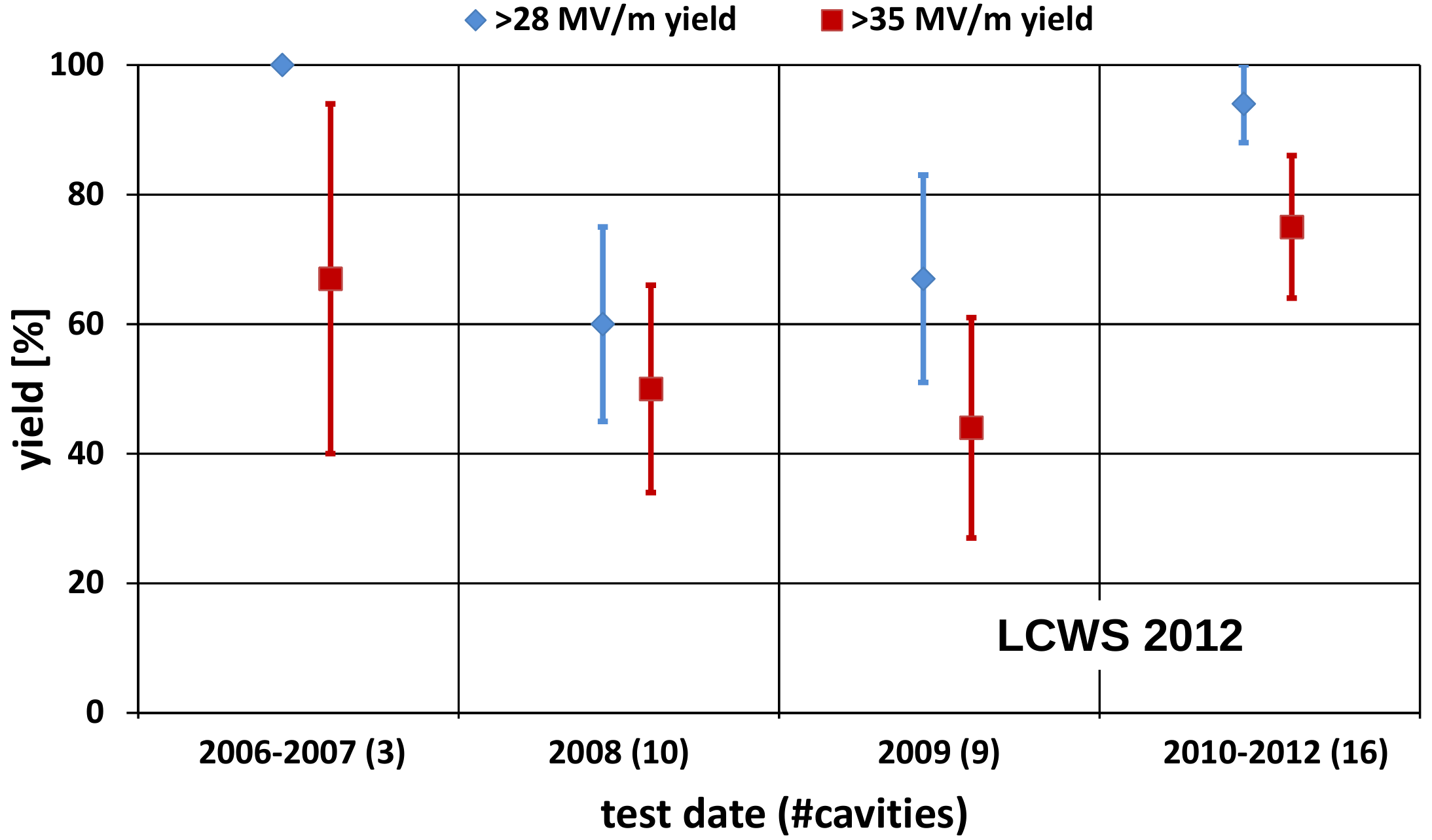}}
\caption{Production yield of ILC-type cavities for two accelerating gradients~\cite{Geng}. The cavities have been processed once or twice following the ILC recipes.}
\label{fig:ILC_cavity_yield}
\end{figure}

While test cavities in the lab have reached accelerating gradients beyond 45~MV/m (figure~\ref{fig:ILC_cavity_gradient}), the challenge is to transfer the production and surface treatment recipes to the industrial scale. Figure~\ref{fig:ILC_cavity_yield} shows the history of the yields of ILC-type cavities. Though the statistics are still low, the yield of 80\% has been reached for the most recently produced cavities. The XFEL cavities that are under production right now will provide the test case for the industrial mass production.

\section{The Detector Challenge}

The ILC physics programme offers a wide range of physics questions and the envisaged detectors need to cope with that. A variety of different final states with many different signatures and event topologies need to be measured with high precision. Other as for the LHC detectors, radiation hardness is not a main driver in the detector design. The best overall event reconstruction will be reached by the ILC detectors following the particle flow concept.

\subsection{The Particle Flow Paradigm}

The particle flow concept is based on the philosophy to reconstruct all particles in a event, especially in multi-jet topologies, using the most precise detector element~\cite{Thomson:2009rp}. At ILC energies, typical events are hadronic final states from Z and W particles. The composition of the jets is typically so that 60\% of the final state particles are charged, around 30\% are photons and around 10\% are neutral long lived hadrons. The charged particles are best reconstructed using the tracking system where the momentum resolution is much better than the energy resolution of the calorimeters. The theoretical best total jet energy resolution is then given by:

\begin{equation}
\sigma^2(E_{jet}) = \omega_{tr}\sigma_{tr}^2 + \omega_{\gamma}\sigma_{\gamma}^2 + \omega_{h^0}\sigma_{h^0}^2 
\end{equation}

where $\omega_i$ are the relative weights of the jet compositions for charged particles (tr), photons ($\gamma$) and neutral hadrons ($h^0$) and $\sigma_i$ are the corresponding detector resolutions. In reality, this ideal resolution is deteriorated by many effects. The biggest ones stem from confusion, where the proper assignment of tracks and calorimeter clusters is not perfect so that contributions from charged particles are added to the neutral energies and vice versa. This is even more difficult when realistic detectors with dead zones, realistic acceptances and assumptions about resolution errors are taken into account. It is a major milestone in the detector developments for the ILC, that realistic detector simulation models that have been benchmarked with prototypes at test beam experiments have been used to develop and challenge the particle flow algorithms. The current simulations show~\cite{Abe:2010aa, Aihara:2010zz} that jet energy resolutions of $\sigma_E/E \sim 3 - 4 \%$ at multi jet events at 500~GeV collision energies are realistic. That would be sufficient, to e.g.\ separate the hadronic decays from Ws and Zs.
\begin{figure}[t]
\centerline{\includegraphics[width=.6\textwidth]{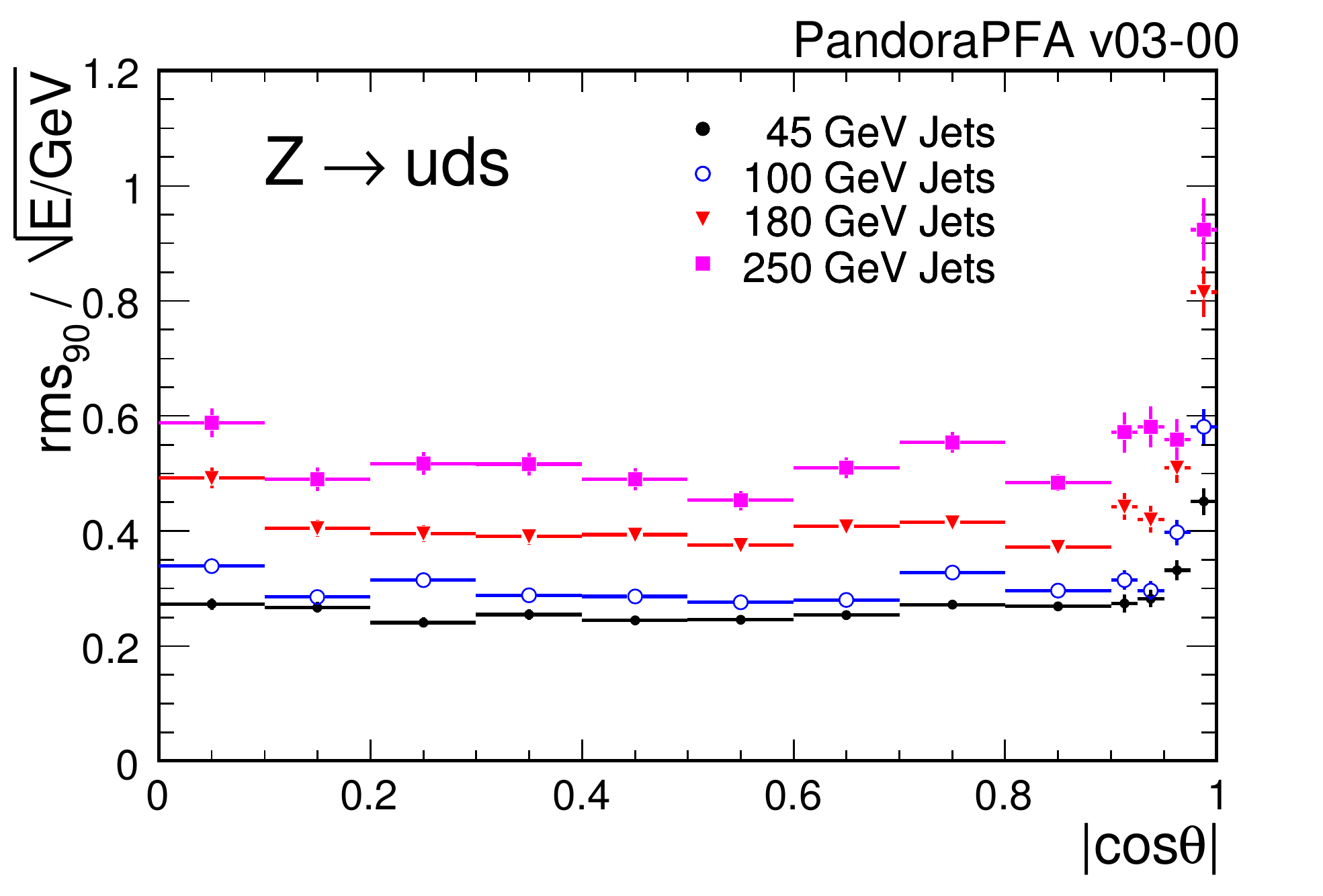}}
\caption{The jet energy resolution at the ILD detector using particle flow algorithms. Shown are the resolutions for four different jet energies~\cite{Abe:2010aa}.}
\label{fig:ILD_pflow}
\end{figure}

\subsection{Detector Concepts}

Two detector concepts are under study for the ILC, ILD~\cite{Abe:2010aa} and SiD~\cite{Aihara:2010zz}, both have been designed with the particle concept in mind. Therefore emphasis has been put on the following design principles:
\begin{description}
\item [Vertexing] Precise vertex detectors that enable precision b- and c-vertex tagging.
\item [Tracking] Precise main tracking systems that are optimised to high resolutions and low material budget.
\item [Granular Calorimeters] A calorimeter system with high granularity for electromagnetic and hadronic interacting particles to keep the confusion in the assignment of tracks and calorimeter clusters as small as possible.
\item [Magnetic Field] High magnetic solenoid fields to suppress beam induced backgrounds and to increase the momentum measurement precision.
\item [Hermeticity] Acceptance down to lowest angles to come as close as possible to a 4$\pi$ detection system.
\end{description}

\begin{figure}[hbt]
\centerline{\includegraphics[width=.7\textwidth]{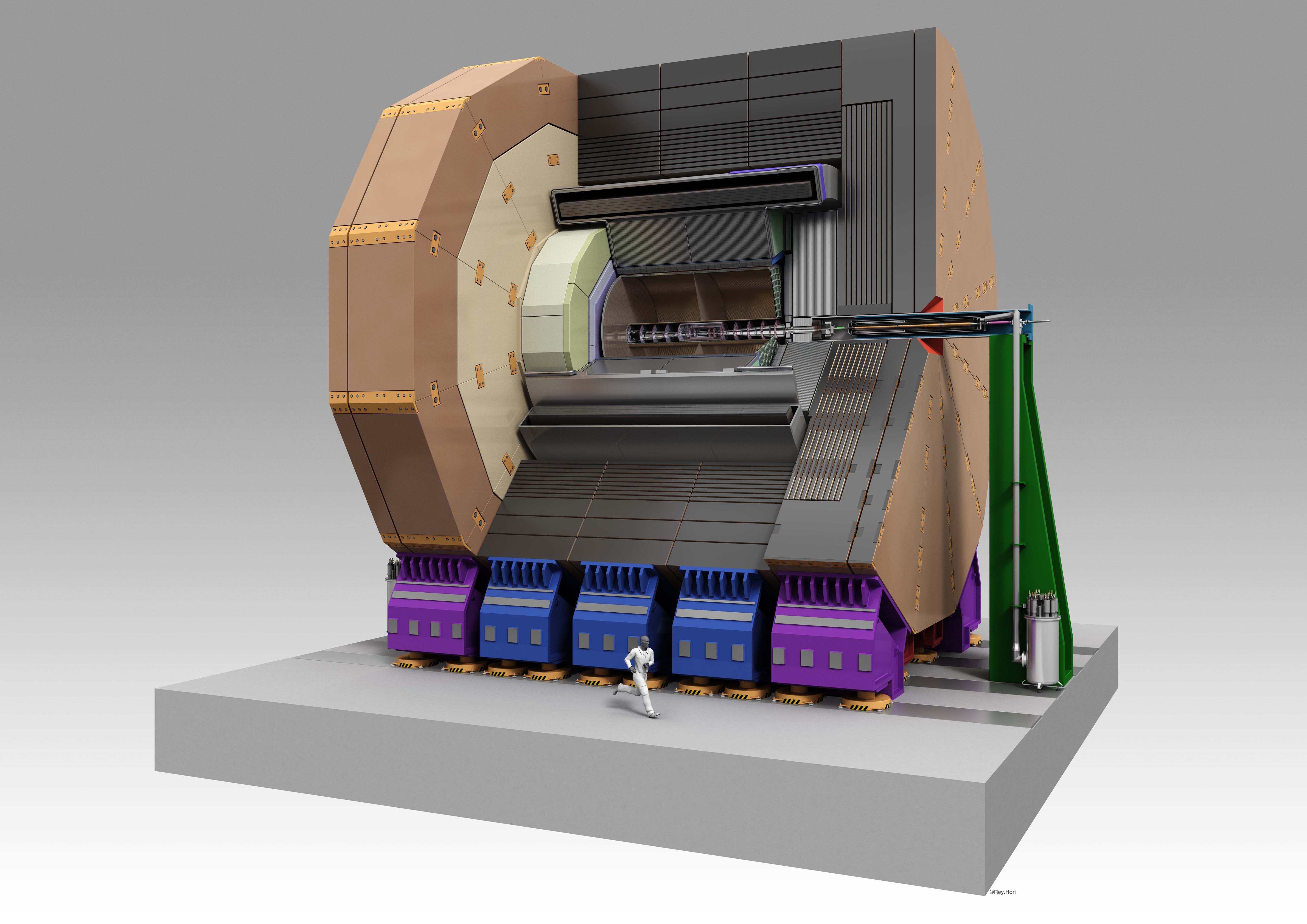}
}
\caption{Artist's view of the ILD detector.}
\label{fig:ILD_design}
\end{figure}

\begin{figure}[hbt]
\centerline{
\includegraphics[width=.7\textwidth]{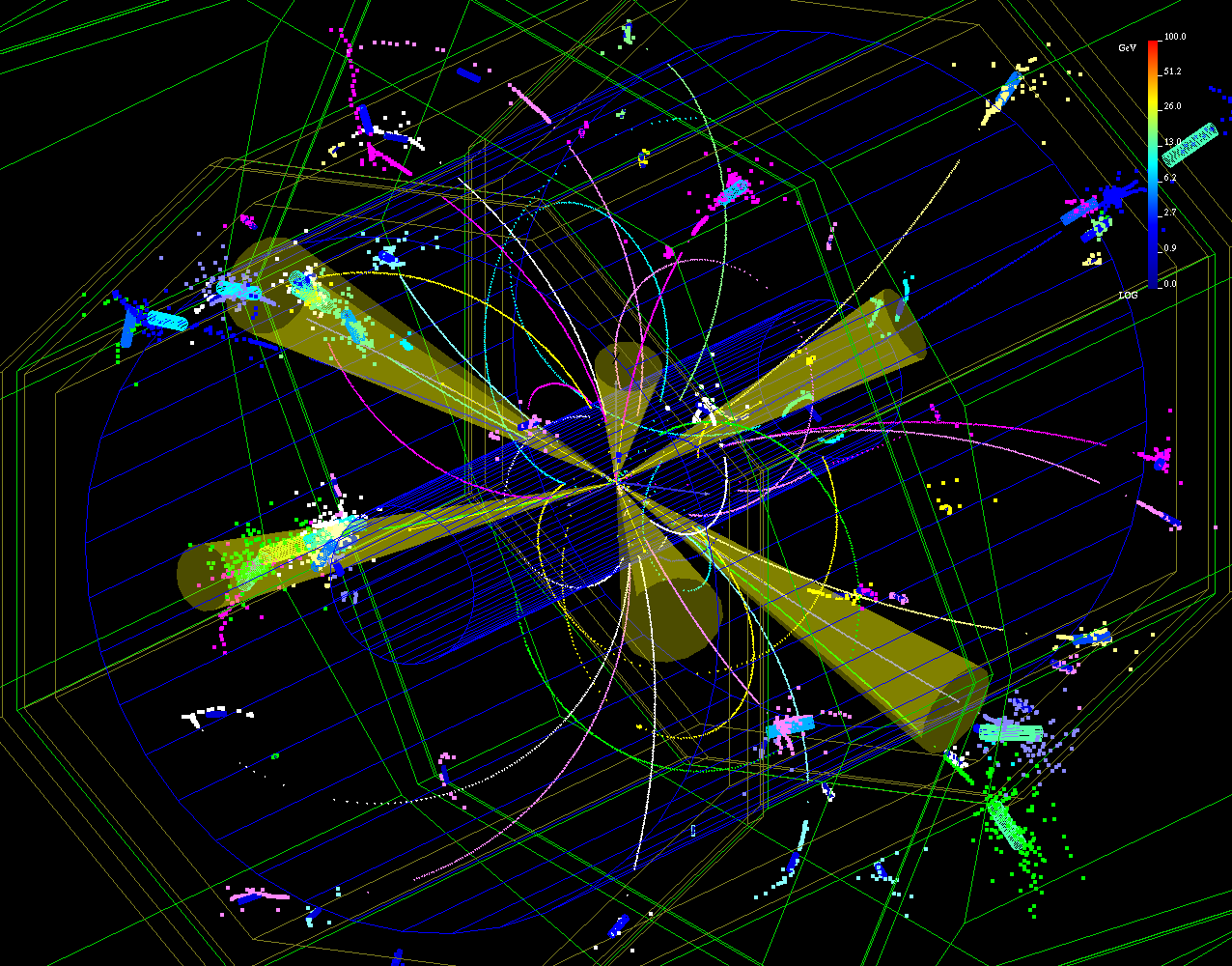}
}
\caption{Simulated e$^+$e$^-$ $\rightarrow$ $t\bar{t}$ 6-jet event at E$_{cm}$ = 500~GeV in the ILD detector that has been reconstructed with particle flow algorithms.}
\label{fig:ILD_tt}
\end{figure}

Figure~\ref{fig:ILD_design} shows and artist's view of the ILD detector. Both concepts have studied benchmark scenarios for the ILC physics programme using full detector Monte Carlo simulations. Realistic beam parameters and overlaid backgrounds have been taken into account for these studies. Figure~\ref{fig:ILD_tt} shows an event display of an e$^+$e$^-$ $\rightarrow$ $t\bar{t}$ 6-jet event at E$_{cm}$ = 500~GeV in the ILD detector that has been reconstructed using particle flow algorithms. The high granularity in the calorimeters allows for the association of calorimeter clusters with charged tracks.

\begin{figure}[t]
\centerline{\includegraphics[width=.4\textwidth]{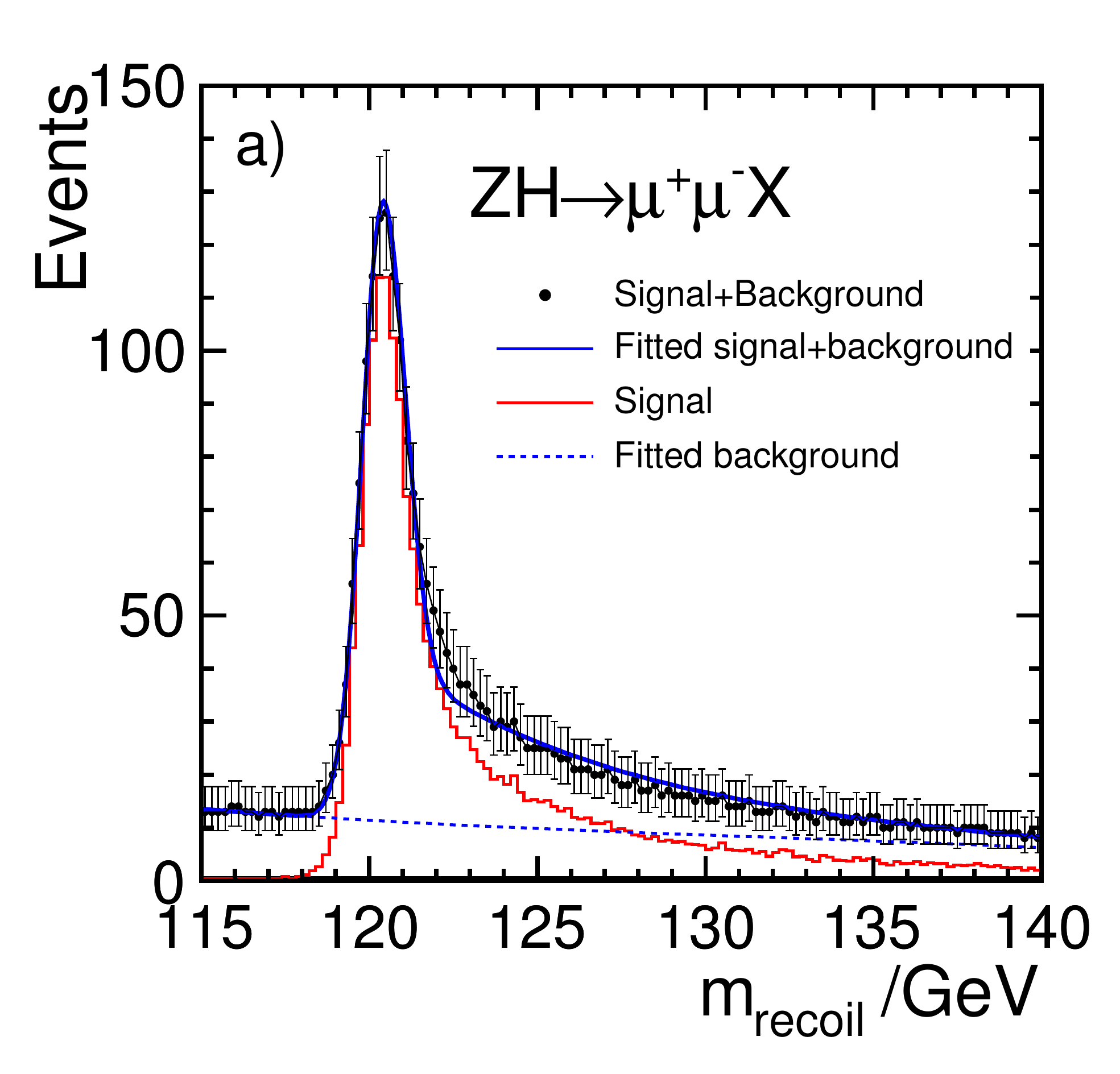}
\includegraphics[width=.4\textwidth]{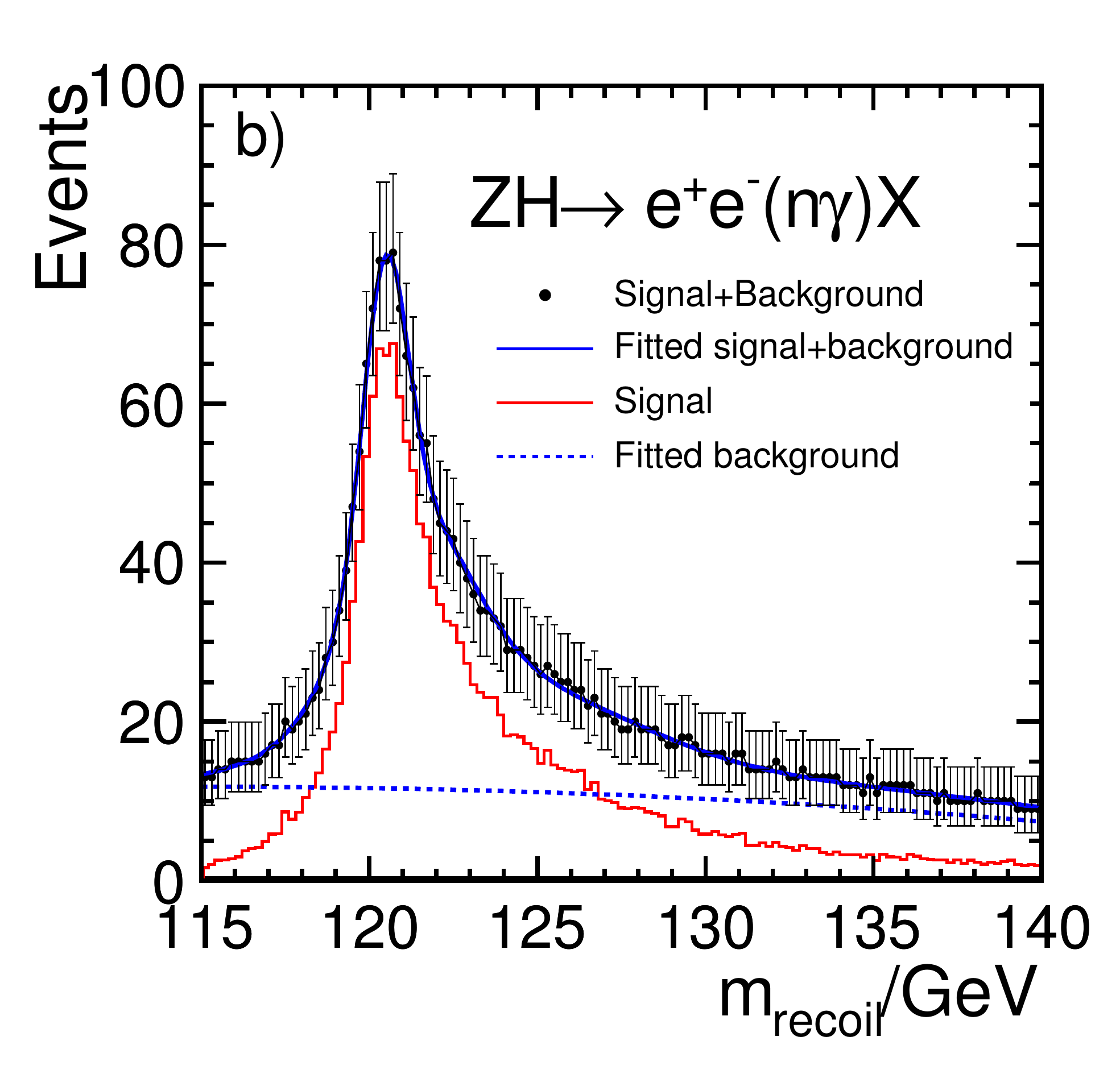}
}
\caption{Results of the model independent analysis of the Higgs-strahlungs process e$^+$e$^-$ $\rightarrow$ HZ for a Higgs mass of 120~GeV in which a) Z $\rightarrow$ $\mu^+\mu^-$ and b) Z $\rightarrow$ e$^+$e$^-$ in the ILD detector~\cite{Abe:2010aa}. In b) a dedicated algorithm to identify bremsstrahlung photons and to include them into the recoil mass calculation has been used.}
\label{fig:ILD_mh}
\end{figure}

Figure~\ref{fig:ILD_mh} shows the results of the reconstruction of the Higgs mass m$_H$ in a model independent way. In the `Higgs-strahlung' process, where the Higgs is produced in association with a Z (e$^+$e$^-$ $\rightarrow$ HZ), the Higgs can be reconstructed by calculating the recoil mass from the well measured Z decay products. No assumptions about the decay of the Higgs need to be done.

\subsection{Two Detectors in Push-Pull Configuration}

\begin{figure}[hbt]
\centerline{\includegraphics[width=.8\textwidth]{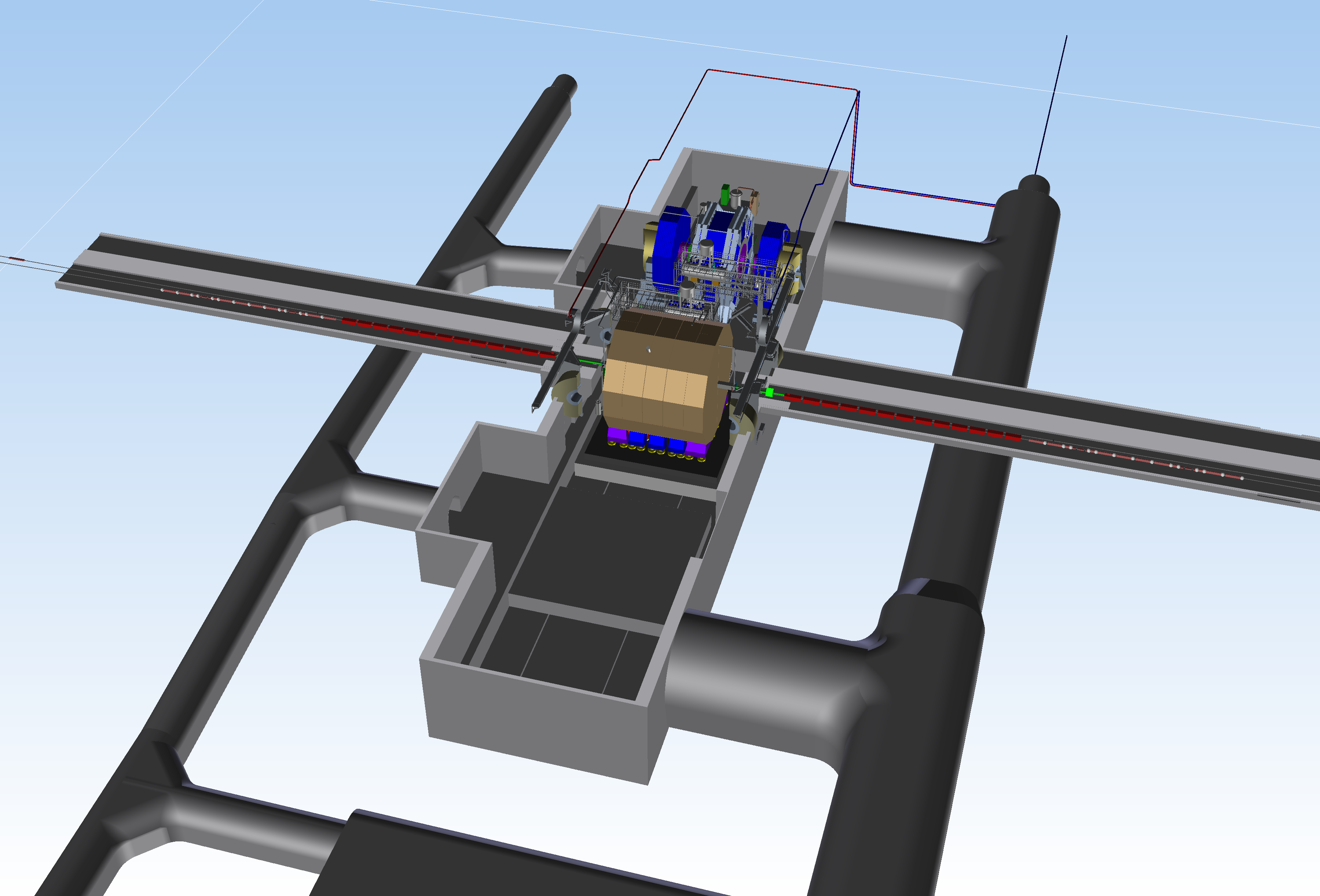}
}
\caption{ILD (on the beam) and SiD (in maintenance position) in the experimental hall in push-pull configuration.}
\label{fig:push-pull}
\end{figure}

Other as in a storage ring, the total integrated luminosity at a linear collider does not scale with the number of interaction regions. Nevertheless, it is crucial to have two detectors with complementary design that are built and exploited by two independent collaborations. In the ILC design, the two detectors would share one beam line with one interaction region. The detectors will be placed on movable concrete platforms that allow a changeover between both in a short time of the order of one day. This is necessary to keep the total integrated luminosity high while frequent switches between both detectors would allow both collaboration a fair sharing of the collision time. It should be avoided to allow one collaboration to collect enough data to make a possible discovery on its own. Figure~\ref{fig:push-pull} shows a possible design of the experimental area with ILD and SiD in push-pull configuration.

\section{Outlook}

The publication of the ILC Technical Design Report (TDR) is envisaged for 2013 and concludes the technical R\&D phase for this most advanced linear collider project. Recently, the CLIC project has published the Conceptual Design Report~\cite{Lebrun:2012hj}. The International Committee for Future Accelerators ICFA has founded the new Linear Collider Collaboration (LCC) that now combines both technical solutions for a linear collider, ILC and CLIC, together with the joint physics and detector efforts into one organisation. This comes in time with the discovery of the Higgs-like boson at the LHC that opens the window into the Terascale physics. 

The ILC is already now the ideal tool to study the new boson with highest precision. The question of whether this new boson is really the Higgs particle that is responsible for the generation of mass, as predicted in the Higgs mechanism of the standard model, will be answered by the ILC results. At the same time the precision measurements will provide access to energy scales that will not be reachable by the LHC. The recent initiative by the Japanese particle physics community to select the realisation of the ILC in Japan as their highest priority has therefore found a worldwide positive echo. The CLIC technology provides a viable option for lepton collisions in the multi-TeV range though on a longer time scale as the ILC.

Thus, the LCC is in the favourable position to participate in both, the next big project in particle physics, whose realisation could start already in this decade in Japan, as well as in the possible post-LHC project at CERN. At the same time the obvious synergies between the ILC and CLIC will be exploited in the accelerator and detector development as well as in the sharpening of the physics programme for lepton physics at the Terascale.

\section{Acknowledgements}
Part of the presented work has been supported by the Commission of the European Communities, grant no.~206711 "ILC-HiGrade" and by the Helmholtz Association, contract HA-101 "Physics at the Terascale". I would like to thank the organisers of the Corfu Summer Institute for a very nice and perfectly organised summer school teaching experience.

\end{document}